\documentstyle[12pt]{article}

\begin{document}

\begin{center}
{\bf The static spacetime relative acceleration for the general free fall
and its possible experimental test}
\end{center}

\begin{center}                             
F. Bunchaft and S. Carneiro
\end{center}

\begin{center}
{\small \it Instituto de F\'{\i}sica, Universidade Federal da Bahia\\
40210-340, Salvador, BA, Brasil.}
\end{center}

\begin{abstract}
Mishra has recently established, using a generic static metric, the
relative local proper-time $3$-acceleration of a test-particle in
one-di\-mensional free fall relative to a static reference frame in any
static spacetime. In this paper, on the grounds of
gravitoelectromagnetism we establish, in a covariant spacetime form, the
relative $4$-acceleration for the general free fall, indicating its
canonical representation with its $3$-space cinematical content. Then we
obtain the relation between this representation and the very known
expression for the relative free fall acceleration in Fermi coordinates.
Taking this into account, it is shown that an experiment with
relativistic beams in a circular accelerator, modelled by Fermi
coordinates, recently proposed by Moliner et al, can test the here
established covariant result and, therefore, can also verify Mishra's
formula. This possibility of experimental verification, besides its
intrinsic importance, can answer a recent inquire by Vigier, related to
his recent proposal of derivation of inertial forces.

\vspace{0.5cm}


\end{abstract}

\section{Introduction}

Mishra$^{[1]}$ has recently presented some relations (the
``transformation law" and the ``addition law" for accelerations) between
cinematical observations on particles in one-dimensional motion in a
general relativistic static spacetime, as performed by 
accelerated observers in this spacetime. From his general relativistic
addition law Mishra obtains, in particular, the formula
$\vec{a}=\gamma^{-2}\vec{g}$ (see equation (11) in $[1]$) for the local
proper-time relative $3$-acceleration $\vec{a}$ (Mishra's ``physical
acceleration") of a relativistic particle in one-dimensional free fall in
any static gravitational field, relative to a preferred non-linear static
reference frame ($\vec{g}$ stands for the acceleration when the
``physical" relative velocity $\vec{v}$ is zero; $\gamma^{-2} = (1-v^2)$
in units in which $c=1$).

More recently J.P.Vigier has pointed out the convenience of performing a
laboratory test to verify this formula. His suggestion comes up in the
course of an extensive article$^{[2]}$ dedicated to expose a possible,
non-machian, solution to the old-age "unsolved mistery in modern physics"
of the origin and nature of inertia. In his proposal, inertial forces
arise from the local interaction of a physical vacuum or ether (in
Dirac's covariant model) with accelerated particle-like solitons piloted
by surrounding wave-packets. For simplicity Vigier considers only the
one-domensional accelerated motion of the solitons in the ether and puts
to work some inertial and accelerated observers/frames. Then the previous
Mishra's formula, says Vigier, ...``plays a crucial role in our
derivation/interpretation of inertia. It evidently proves that the
acceleration due to a pseudo-force (inertial force) and that due to the
force of gravity are both decreased by a factor $(1-v^2/c^2)$..."

In this paper, we situate the problem in an enlarged context, following
the gravitoelectromagnetism (GEM) formalism as outlined by Bini et
al$^{[3]}$. These authors have, in particular, re-obtained, in a
coordinate-free form, all the results of Mishra.
 
Our first step is to extend, in a covariant spacetime form, Mishra's
formula to the case of the static general free fall. Then we represent
the obtained $4$-covariant formula in a canonical form in the observers
computational $3$-space and proper-time. Further on, introducing
observer-adapted Fermi coordinates (FC) we obtain the relations between
the canonical representation and the well known general free-fall
relative Fermi-coordinates acceleration $\vec{a}_{FC}$.

As we will see, $\vec{a}_{FC}$ does not coincide with the relative
acceleration in the canonical representation and does not contain
Mishra's one-dimensional expression. At contrary, the measurable physics
which stems from $\vec{a}_{FC}$ for the one-dimensional free fall is
``strange" and radically different from that one which comes from Mishra's law.

Indeed, for $\hat{v}_{FC}=\hat{g}$ it will be $\vec{a}_{FC} =
d\vec{v}_{FC}/dx^0_{FC} = k\vec{g}$, with $k=(1-2v_{FC}^2)$. Thus, only
for $v_{FC}<2^{-1/2}$ it will be $k>0$. For $v_{FC}=2^{-1/2}$, the
particle will follow a FC-uniform movement and, when $v_{FC}>2^{-1/2}$,
it will be $k<0$. Clearly, this kind of ``anomalies" is absent from
Mishra's expression\footnote{Jaffe and Shapiro$^{[4]}$ have already
obtained the same ``anomalies" for the spherically symmetric
gravitational field in Schwarzchild coordinates.}.

The obtainment of a connection between the canonical relative
$3$-accelera\-tion representation of the GEM spacetime formula and
Fermi's relative $3$-acceleration allows us to reconsider an experiment
recently proposed by Moliner et al$^{[5]}$ with almost
horizontal relativistic particles in a circular accelerator in order to
test $\vec{a}_{FC}$, with Fermi coordinates modelling the experiment. In
view of the previously obtained relations between the GEM formulae and
$\vec{a}_{FC}$, we show that the realm
of the experiment can be considerably enlarged to test also the GEM
expressions and its theoretical implications, in particular Mishra's
formula. This, besides to be a possible test of general relativity in a
terrestrial experiment involving relativistic
massive particles, would also be significant to estimate Vigier's
derivation of inertial forces.

\section{The GEM formula for the general free fall}

{\bf 2.1.} Let us initiate with a brief recall on Bini et al exposition.

Spatial gravitational forces modelled after the electromagnetic
$4$-force, that is, ``gravitoelectromagnetic forces", rely on the
splitting of spacetime by means of a congruence of test-observers $(u)$.
The decomposition of each tangent space into a local
direction along the $4$-velocity vector field $u$ of $(u)$ and its
orthogonal complement, the local instantaneous $u$-rest space $LRS_u$,
induces a corresponding coordinate-free decomposition of all spacetime
tensors and tensor equations, leading to spatial spacetime tensor fields
(any contraction with $u$ gives zero) and spatial equations which
represent them, i.e., which ``measure" them. This decomposition is
accomplished by $T(u)$ and $P(u)$, the operators of temporal projection
and of spatial projection into $LRS_u$, respectively, being
$P(u)X=X+u[u\cdot X]$. These operators may be identified with suitable
mixed second rank tensors acting by contraction.

Through $P(u)$ there are also introduced spatially projected differential
operators: so, from the spacetime covariant derivation operator
$\;^4\nabla$, it arises the spatial covariant derivative
$\nabla(u)=P(u)\;^4\nabla$, the spatial Fermi-Walker derivative
$\nabla_{fw}(u)=P(u)\;^4\nabla_u$
(which, for spatial tensor fields coincides with the spacetime
Fermi-Walker derivative along $u$) etc.

Considering now a test-particle $(U)$ with $4$-velocity $U$, the
orthogonal decomposition of $U$ w.r.t $u$ defines the relative velocity
$v(U,u)$ of $(U)$ w.r.t to $(u)$ and the associated gamma factor, i.e.,

\begin{equation}
\label{1}
U=\gamma(U,u)[u+v(U,u)]
\end{equation}

\noindent where

\begin{equation}
\label{2}
\gamma(U,u)=[1-v(U,u)\cdot v(U,u)]^{-\frac{1}{2}}
\end{equation}

\noindent (standing the dot for the inner product of spacetime vectors).
$v(U,u)$ is $u$-spatial, since it is the rescaled $u$-spatial projection
of $U$, i.e.,

\begin{equation}
\label{3}
v(U,u)=\gamma(U,u)^{-1}P(u)U
\end{equation}

Now let $a(U)=\nabla_UU$ be the $4$-acceleration of $(U)$ and let
$a(U)=\tilde{f}(U)$ be the equation of motion for $U$, being 
$\tilde{f}(U)$ the $4$-force per unit mass on the test-particle $U$. Then
the ortogonal decomposition and the spatial projection
of the spacetime tensors and of the equation of motion lead to

\begin{equation}
\label{4}
A(U,u)=\tilde{F}(U,u)
\end{equation}

\noindent where

\begin{equation}
\label{5}
A(U,u)=\gamma(U,u)^{-1}P(u)a(U)
\end{equation}

\noindent (and equivalently for $\tilde{F}(U,u)$ and $\tilde{f}(U)$). 

Expressing $A(U,u)$ in terms of the relative momentum per unit mass,
$\tilde{p}(U,u)=\gamma(U,u)v(U,u)$, introducing the composite projection
map defined by $P(u,U,u)=P(u)P(U)P(u)$, which is an automorphism of
$LRS_u$, and defining

\begin{equation}
\label{6}
a_{fw}(U,u)=\gamma(U,u)^{-1}P(u)\nabla_Uv(U,u)
\end{equation}

\noindent as the Fermi-Walker relative acceleration of $U$ w.r.t. $u$,
Bini at al have derived the expression

\begin{equation}
\label{7}
A(U,u)=-\gamma(U,u)[g(u)+H_{fw}(u)v(U,u)]+\gamma(U,u)P(u,U,u)a_{fw}(U,u)
\end{equation}

Here

\begin{equation}
\label{8}
g(u)=-a(u)
\end{equation}

\begin{equation}
\label{9}
H_{fw}(u)=-\nabla(u)u=\omega(u)-\theta(u)
\end{equation}

\noindent being $\omega(u)$ and $\theta(u)$, respectively, the vorticity
(rotation) and expansion tensors of the observers family $(u)$, and
being implied a contraction between $H_{fw}(u)$ and $v(U,u)$.

Now, $a_{fw}(U,u)$ can be rewritten as 

\begin{equation}
\label{10}
a_{fw}(U,u)=\frac{D_{fw}(U,u)}{d\tau_{U,u}}v(U,u)
\end{equation}

\noindent where $D_{fw}(U,u)/d\tau_{U,u}$ is the Fermi-Walker total
spatial covariant derivative along the world-line $U$ expressed in terms
of a parametrization corresponding to the sequence
of differential proper-times of the observers $(u)$ along the world-line
$U$, so that

\begin{equation}
\label{10'}
\frac{d\tau_{U,u}}{d\tau_U} = \gamma(U,u)
\end{equation}

\noindent where $d\tau_U$ corresponds to the $U$-proper-time parametrization.

Besides, it can also be obtained 

\begin{equation}
\label{11}
\frac{D_{fw}(U,u)}{d\tau_{U,u}}\tilde{p}(U,u)=\tilde{F}^G_{fw}(U,u)+\tilde{F}
(U,u)
\end{equation}

\noindent being

\begin{equation}
\label{12}
\tilde{F}^G_{fw}(U,u)=\gamma(U,u)[g(u)+H_{fw}(u)v(U,u)]
\end{equation}

\noindent This allows us to interpret $\tilde{F}^G_{fw}(U,u)$ as the
relative spatial gravitational force on $U$ w.r.t. $u$, being $g(u)$ the
gravitoelectric vector-force field and $H_{fw}(U,u)$ the gravitomagnetic
one. (For a Minkowky spacetime in which 
$(u)$ is an inertial observer, $\tilde{F}^G_{fw}(U,u)$ is zero.)

\vspace{0.5cm}

\noindent {\bf 2.2.} We will now extend Misrha's result, obtaining the
expression for the relative $4$-acceleration in the case of the general
free fall.

Let us begin observing that, in our case, the test-particle $U$ is free,
so the $4$-acceleration $a(U)$ is null. Thus we have $A(U,u)=0$. 

Besides, the spacetime is static and $(u)$ is the prefered local
reference frame, that is, $u$ is the direction of a time-like Killing
vector field, which implies that $\theta(u)=\omega(u)=0$. So,
$H_{fw}(u)=0$ and (\ref{7}) reduces to 

\begin{equation}
\label{13}
P(u)P(U)P(u)a_{fw}(U,u)=g(u)
\end{equation}

Taking now into account that, by definition, $a_{fw}(U,u)$ and $v(U,u)$
are $u$-spatial and using the orthogonal decomposition (\ref{1}) of $U$,
the succesive projections (\ref{13}) lead to

\begin{equation}
\label{14}
a_{fw}(U,u)+\gamma^2[a_{fw}(U,u)\cdot v(U,u)]v(U,u)=g(u)
\end{equation}

Now, multiplying this equation by $v(U,u)$ and considering the definition
(\ref{2}) of $\gamma(U,u)$, it is obtained

\begin{equation}
\label{15}
\gamma^2[a_{fw}(U,u)\cdot v(U,u)]v(U,u)=g(u)\cdot v(U,u)
\end{equation}

\noindent or, finally,

\begin{equation}
\label{16}
a_{fw}(U,u)=g(u)-[g(u)\cdot v(U,u)]v(U,u)
\end{equation}

\noindent which is the $4$-covariant GEM expression for the general free
fall in a static spacetime\footnote{This result can also be obtained
directly by using the expression of $[P(u)P(U)P(u)]$ as given in line 1
of Table 1 of reference $[3]$. Or, alternatively, by
solving (\ref{13}) for $a_{fw}(U,u)$, using the expression of
$[P(u)P(U)P(u)]^{-1}$ as given in line 2 of the table, as suggested by an
anonimous referee.}.

It is immediate to verify that, for $\hat{\nu}(U,u) = \hat{g}(u)$, it results

\begin{equation}
\label{17}
a_{fw}(U,u)=\gamma^{-2}g(u),
\end{equation}

\noindent which is the spacetime covariant expression of Mishra's result
for the one-dimensional free fall.

\section{The canonical representation}

Equation (\ref{16}) as well as (\ref{17}) previously derived are general
covariant equations for the free-fall $4$-acceleration relative to a
preferred static reference frame where the observers constitute a
time-like Killing vector field, always present 
in some open submanifold of any stationary spacetime. Since our generic
spacetime metric is not only stationary but, more than that, static, the
observers local rest space constitutes a spatial-like slicing orthogonal
to the observers spacetime threading.

Thus we can complete the spacetime threading to an observer-adapted
frame, that is, any frame $\{e_{\alpha}\}$, $\alpha=0,1,2,3$, such that
$e_0$ is along the observer $4$-velocity $u$ and the spatial frame
$\{e_i\}$, $i=1,2,3$, spans the local rest space
at each point along $u$. Besides, the spatial frame is Fermi-Walker
transported along $u$, which assures that it remains orthogonal (and even
orthonormal if we choose so). Finally we take for our present purposes
the frame basis to be a coordinate-one, being $\{x^{\alpha}\}$ the local
coordinates adapted to the frame.

Under these conditions, the general theory shows that the algebra of
stationary spatial tensors is isomorphic to the tensor algebra of the
computational $3$-space equipped with the time-independent projected
spatial metric $\gamma_{ij}$ expressable in the
local adapted coordinates$^{[3,6]}$.

Furthermore, since the $g_{\alpha\beta}$ metric is static, any
spatial-like orthogonal slice can be taken as the computational space and
$\gamma_{ij}=g_{ij}$. Besides, the spatial operators of static spatial
fields reduce to the correspondent operators defined with respect to
$\gamma_{ij}$.
From hereafter a non-linear static reference frame will be always
considered equipped with the above defined structures.

Now, returning to the $4$-covariant GEM equation referred to such a
static reference frame, $g(u) = -\nabla_uu$ and, since $u$ is a unit
vector, $g(u)$ is spatial and, by definition, $a_{fw}(U,u)$ and $v(U,u)$
also are. Then the projected equation can be noted in the $3$-space vector
notation

\begin{equation}
\label{18}
\vec{\tilde{a}}=\vec{g}-\vec{g}\cdot \vec{\tilde{v}}\vec{\tilde{v}}
\end{equation}

\noindent where $a_{fw}(U,u)=(0,\vec{\tilde{a}})$, $g(u)=(0,\vec{g})$,
$v(U,u)=(0,\vec{\tilde{v}})$ and the spatial inner product can be
considered as arising from

\begin{equation}
\label{19}
X\cdot_uY=P(u)_{\alpha\beta}X^{\alpha}Y^{\beta}=g_{ij}X^iY^j=\gamma_{ij}x^iy^j
\end{equation}

\noindent for any pair of spatial vector fields $X=(0,\vec{x})$ and
$Y=(0,\vec{y})$. The projected equation (\ref{18}), for its naturalness,
will be said the {\it canonical representation} of the covariant GEM
expression (\ref{16}).

The generic relation between the projected variables $\vec{\tilde{v}}$
and $\vec{\tilde{a}}$ in the canonical representation and the relative
$3$-geometric cinematical variables is also given by the general
theory$^{[3,6]}$, but it will be worthwhile for
our immediate purposes to unfold it here directly.

Then let ${x^{\alpha}_U}$ be the coordinates of some particle world-line
with $4$-velocity $U$. So
${\cal U}^{\alpha}=dx^{\alpha}_U/dx^0=\dot{x}^{\alpha}_U$ will denote the
coordinate-components of the coordinate-velocity of $U$ world-line, whose
components are

\begin{equation}
\label{20}
U^{\alpha}=\frac{dx^{\alpha}_U}{d\tau_U}=\dot{x}^{\alpha}_U\left(\frac{dx^0_U}
{d\tau_U}\right)=\Gamma_{x^{\alpha}}(U,u)\dot{x}^{\alpha}_U
\end{equation}

\noindent being

\begin{equation}
\label{21}
\frac{dx^0_U}{d\tau_U}=(g_{\alpha\beta}\dot{x}^{\alpha}_U\dot{x}^{\beta}_U)^
{-\frac{1}{2}}=({\cal U}_{\alpha}{\cal 
U}^{\alpha})^{-\frac{1}{2}}\equiv\Gamma_{x^{\alpha}}(U,u)
\end{equation}

\noindent the $x^{\alpha}$-coordinate Lorentz factor.

Let us now denote $v^i$ the coordinates of a $3$-vector $\vec{v}$ defined by

\begin{equation}
\label{22}
v^i=\frac{dx^i_U}{d\tau_{U,u}}=\dot{x}^i_U g_{00}^{-1/2}
\end{equation}

\noindent since $dx^0_U/d\tau_{U,u}=g_{00}^{-1/2}$ in the static metric
expressed in $x^{\alpha}$-coordinates.

Note that this definition implies that

\begin{equation}
\label{23}
v^2=\left(\frac{dl}{d\tau_{U,u}}\right)^2
\end{equation}

\noindent where $dl^2=g_{ij}\;dx^i_U\;dx^j_U$. So $\vec{v}$ is the
$3$-velocity of the $U$-particle measured by the $u$-observer at rest in
the same place as the $U$-particle, in the proper-time of the observer,
to be called hereafter the local proper-velocity of the particle $U$.

Then, denoting $\gamma^*(U,u)=[1-v^2]^{-1/2}$, we have

\begin{equation}
\label{24}
\Gamma_{x^{\alpha}}(U,u)=g_{00}^{-1/2}[1-g_{ij}v^iv^j]^{-\frac{1}{2}}=g_{00}^
{-1/2}\gamma^*(U,u)
\end{equation}

\noindent Thus $\gamma^*(U,u)=\gamma(U,u)=d\tau_{U,u}/d\tau_U$, as
defined in (\ref{10'}), and

\begin{equation}
\label{25}
U^i=\gamma(U,u)v^i
\end{equation}

\begin{equation}
\label{26}
U^0=\gamma(U,u)g_{00}^{-1/2}
\end{equation}

Therefore

\begin{equation}
\label{27}
v^i=\gamma^{-1}(U,u)U^i=\gamma^{-1}(U,u)[P(u)U]^i
\end{equation}

\noindent which, by definition of $v(U,u)$ (see (\ref{3})), leads to
$v^i=v^i(U,u)=(0,\vec{\tilde{v}})^i$, thus identifing $\vec{\tilde{v}}$
with $\vec{v}$, that is, with the local $3$-proper-velocity of the
particle $U$ relative to the stationary observer $u$.

Let us consider now the cinematical meaning of the spatial projection
$\vec{\tilde{a}}$ of $a_{fw}(U,u)$ in the spatial geometry associated to
the static reference frame. From definition (\ref{10}) it comes

\begin{eqnarray}
\label{28}
a_{fw}(U,u)=P(u)\frac{D}{d\tau_{U,u}}[\gamma^{-1}(U,u)P(u)U]=P(u)\frac{D}
{d\tau_{U,u}}(0,\vec{\tilde{v}})=(0,\vec{\tilde{a}})
\end{eqnarray}

\noindent so that we can identify the projected acceleration
$\vec{\tilde{a}}$ with

\begin{equation}
\label{29}
(\vec{a})^i = \left( 
\frac{^3D\vec{v}}{d\tau_{U,u}} \right)^i =
\left( \frac{d\vec{v}}{d\tau_{U,u}} \right)^i +
\Gamma^i_{jk}v^jv^k
\end{equation}

\noindent where $\Gamma^i_{jk}$ are the components of the Riemannian
connection associated to the $3$-metric $g_{jk}$. That is, we can
identify $\vec{\tilde{a}}$ with $\vec{a}$, the local
$3$-proper-acceleration of the particle $U$ relative to the stationary
observer $u$.

So, we have shown that the canonical representation of the covariant GEM
formula can be identified with the expression

\begin{equation}
\label{30}
\vec{a}=\vec{g}-\vec{g}\cdot\vec{v}\;\vec{v}
\end{equation}

\noindent where $\vec{a}$ and $\vec{v}$ have now a precise $3$-geometric
cinematical content.

Let us observe that an expression like (\ref{30}) has been recently 
presented by Mould$^{[7]}$ for uniformly accelerated frames in the flat
spacetime, starting from the specific metric to obtain the
coordinate-acceleration and changing then to the suitable local
proper-observers (see equation (8.45) in $[7]$). So, the present GEM
derivation can be said to extend Mould's expression to any static
spacetime and Mishra's result to the general free fall.

\section{The GEM formulae and the cinematical variables in Fermi coordinates}

Let us consider Fermi-coordinates adapted to a stationary observer
modelled by a world-line $u$ with covariant $4$-acceleration
$a(u)=-g(u)$. As these coordinates are suitable to model terrestrial
experiments like the one considered in the next section, let us now try
to connect the GEM formulae, through the canonical representation
previously derived, with the cinematical variables in these particular
coordinates.

For this purpose we need to establish the suitable relations between the
local cinematical variables, in the canonical representation, and the
adapted Fermi-coordinates cinematical variables. (Clearly, in this
context, a preliminary assumption must be the
Fermi-Walker transport of the Fermi spatial frame, in order to assure
orthonormality and ``non-rotation" of the frame.)

Locally, in Fermi coordinates, the spacetime metric will be, as it is
well known (see equation (13.71) in $[8]$)

\begin{equation}
\label{31}
ds^2=(1-2g_ix_{FC}^i)(dx^0_{FC})^2 - \delta_{ij}dx^i_{FC}dx^j_{FC}+O
\left(|x^i_{FC}|^2\right)dx^{\alpha}_{FC}dx^{\beta}_{FC}
\end{equation}

\noindent being $\alpha,\beta=0,1,2,3$; $i,j=1,2,3$; $g(u)=(0,\vec{g})$;
$g_i=(\vec{g})_i$. Stationarity implies that $\vec{g}$ does not depend on
$x^0_{FC}$.

This means that our observer has been immersed in a family of local
stationary observers $(u)$, whose local rest spaces integrate to a
spatial-like hypersurface which is locally flat at this order of
approximation, but whose coordinate clocks are all paced by our observer
at the origin. Let us recall that, in the previous GEM equations, all the
local observers at rest in the frame reparametrize any geodesic
world-line $U$ of a particle in free fall by their own local proper-time
$\tau_{U,u}$ (not by the coordinate-time $x^0$ nor by the proper-time
$\tau_U$ of $U$).

Now, let us apply (\ref{22}) and (\ref{29}) to the adapted Fermi
coordinates. With the notation $v^i_{FC}=\dot{x}^i$,
$a^i_{FC}=\ddot{x}^i$, one has

\begin{equation}
\label{32}
\vec{v}=g_{00}^{-1/2}\dot{\vec{x}}=(1-2g_ix^i_{FC})^{-\frac{1}{2}}\vec{v}_{FC}
\end{equation}

\begin{equation}
\label{33}
\vec{a}=\frac{^3D\vec{v}}{d\tau_{U,u}}=\frac{d\vec{v}}{d\tau_{U,u}}
\end{equation}

\noindent since the spatial $3$-metric $g_{ij}=\delta_{ij}$ is flat.
Then, from $d\tau_{U,u}=g_{00}^{1/2}dx_{FC}^0$ it will be

\begin{equation}
\label{34}
\vec{a}=g_{00}^{-1/2}\frac{d}{dx_{FC}^0}(g_{00}^{-1/2}\vec{v}_{FC})=
(1-2g_ix_{FC}^i)^{-1}\vec{a}_{FC}+(1-2g_ix_{FC}^i)^{-2}g_jv^j_{FC}\vec{v}_{FC}
\end{equation}

For the observer at the origin we have

\begin{equation}
\label{35}
\vec{v}=\vec{v}_{FC}
\end{equation}

\begin{equation}
\label{36}
\vec{a}=\vec{a}_{FC}+\vec{g}\cdot \vec{v}_{FC}\;\vec{v}_{FC}
\end{equation}

\noindent which establishes the relations between the canonical
cinematical variables and the Fermi-coordinate ones.

On the other hand, $\vec{a}$ is expressed by (\ref{30}),
so that, in view of (\ref{35}), 

\begin{equation}
\label{38}
\vec{a}=\vec{g}-\vec{g}\cdot \vec{v}_{FC}\;\vec{v}_{FC}
\end{equation}

The consistence of the pair of equations (\ref{36}) and (\ref{38}) for
$\vec{a}$ can be verified by obtaining from them the very well known expression

\begin{equation}
\label{39}
\vec{a}_{FC}=\vec{g}-2\vec{g}\cdot \vec{v}_{FC}\;\vec{v}_{FC}
\end{equation}

\noindent for the Fermi coordinate-acceleration, which is usually
directly obtained by a very distinct derivation (see equation (13.75) of
$[8]$).

Note the essential theoretical difference between $\vec{a}$ and
$\vec{a}_{FC}$: $\vec{a}$ is constructed in a coordinate-free manner and
so is invariant; the change of the temporal parametrization (from the
observers proper-time to the Fermi coordinate-time) of the test-particle
world line when we go from the canonical representation to the Fermi
coordinates produces the variation
$\vec{a}-\vec{a}_{FC}=\vec{g}\cdot\vec{v}_{FC}\;\vec{v}_{FC}\neq0$, when
$\vec{g}\cdot\vec{v}_{FC}\neq0$ (or $g_ix^i_{FC}\neq 0$), since clocks
at different heights beat at different rates (otherwise,
$\vec{a}=\vec{a}_{FC}=\vec{g}$). (In what concerns spatial coordinates,
let us recall that the Fermi ones measure proper-distances, in this order
of approximation.)

Clearly, from the expression for $\vec{a}_{FC}$, changing from
coordinate-time to local proper-time, it would be possible to derive
$\vec{a}$, but this would hidden the canonical content of $\vec{a}$ and
of equation (\ref{30}) as the canonical $3$-space representations of the
spacetime covariant Fermi-Walker relative acceleration $a_{fw}(U,u)$ and
of the covariant equation (\ref{16}), respectively - that is, the
covariant content of the problem.

\section{A possible test for the GEM expressions}

Moliner et al$^{[5]}$ have recently suggested a possible way to test the
Fermi coordinate-acceleration expression. In their suggestion, such a
test is to be performed in a circular accelerator where a charged
particle moves under the influence of suitable
electric and magnetic fields, besides, of course, the Earth gravitational one.

Then, equation (\ref{39}) is extended to give

\begin{equation}
\label{40}
\vec{a}_{FC} = \vec{g} - 2\vec{g} \cdot \vec{v}_{FC} \vec{v}_{FC}+\frac{e}
{m\Gamma_{FC}(U,u)} (\vec{E} + \vec{v}_{FC} \times \vec{H} - \vec{E} \cdot 
\vec{v}_{FC} \vec{v}_{FC})
\end{equation}

\noindent where $\vec{E}$ and $\vec{H}$ are the electric and magnetic
fields, $e$ is the charge, $m$ is the mass and
$\Gamma_{FC}(U,u)=(1-v_{FC}^2)^{-1/2}$ is the Fermi-coordinate Lorentz factor.

Defining

\begin{equation}
\label{41}
\vec{E}_p = \vec{E} + \frac{m\Gamma_{FC}(U,u)}{e} \vec{g}
\end{equation}

\noindent equation (\ref{40}) can be rewritten in the form

\begin{equation}
\label{42}
\vec{a}_{FC} = \frac{e}{m\Gamma_{FC}(U,u)} (\vec{E}_p + \vec{v}_{FC} \times 
\vec{H} - \vec{E}_p \cdot \vec{v}_{FC} \vec{v}_{FC})-\vec{g} \cdot 
\vec{v}_{FC} \vec{v}_{FC}
\end{equation}

\noindent being $-(m\Gamma_{FC}(U,u)/e)\vec{g}$ the electric field
necessary to prevent the particle falling down.

Taking for $\vec{H}$ a uniform magnetic field and for $\vec{E}_p$ a
periodic electric field, both in $\hat{g}$-direction, the term
$-\vec{g}\cdot\vec{v}_{FC}\;\vec{v}_{FC}$ leads to a measurable
horizontal drift of the trajectory, in consequence of a ressonance effect
arising from making the frequency of $\vec{E}_p$ equal to the Larmor
frequency of the particle in the magnetic field.

If $\vec{E}_p=0$, the particle movement will be strictly horizontal and
circular, since then $\vec{E}$ would only prevent the following down.
Switching the additional electric field $\vec{E}_p$, a vertical periodic
component is summed up to the movement and the ressonance effect can
arise. The final result is that the trajectory of the particle projected
in the horizontal plane is now a drifting circle. The final vertical
velocity comes from $\vec{E}_p$ and from the gravitational acceleration
arising from
the term $\vec{g}\cdot\vec{v}_{FC}\;\vec{v}_{FC}$. This velocity also
contributes to the horizontal component of the particle acceleration
through the horizontal part of this same term.

This way it seems possible to verify experimentaly the Fermi
coordinate-acceleration. Remembering that $\vec{a}_{FC}$ can be
correlated, through the canonical representation, to the GEM formula for
the general free fall and that the latter contains Mishra's result, we
are led to reconsider the Moliner et al experiment as a possible way to
test also both these expressions.

Surely, the following question can be posed: using a circular accelerator
in Moliner et al experiment, why is that we are obliged to work (and
measure) $\vec{a}_{FC}$ instead of working and measuring directly
$\vec{a}$? The situation is as follows:

\begin{description}

\item a) in the canonical representation, the trajectory of the particle
is time-parametrised by the proper-time clock of the observer at rest at
the spatial position of the particle. At each position, $\vec{a}$ is
local  (locally measured), since it refers to an arbitrary small spatial
neighbourhood during an arbitrary small interval of time. From the local
point of view, all the observers are mutually independent, none is
preferred and each one measures $\vec{a}$ at its position (in an
arbitrary small neighbourhood). But one cannot characterize a circular
trajectory in such a neighbourhood;

\item b) to characterize (to measure) a circular trajectory in the
accelerator one must refer to fixed, finite, spatial parameters (e.g.,
the radius, cartesian coordinates etc), which cannot be captured
(measured) in such arbitrary small neighbourhoods, so one will
necessarily be led to define non-local simultaneity (i.e., Einstein's
sincronization plus a common rate of the finitelly separated clocks);

\item c) if the accelerator experiment deals only with strictly
horizontal circular trajectories, then all the proper-time clocks at rest
at the same height are equivalent (they beat at the same rate), so the
finitelly separated clocks are already naturally coordinated and there is
no problem: since then
$g_ix^i_{FC}=0$, i.e., $\vec{g}\cdot\vec{v}_{FC}=0$, one will have
(measure) strictly $\vec{a}=\vec{a}_{FC}=\vec{g}$ and no drift. But this
implies only the first half of Moliner et al experiment; 

\item d) to reach the entire scope of the experiment, the trajectory of
the test-particle, by necessity, must be strictly non-horizontal: in
fact, it is made to oscilate vertically w.r.t. the horizontal
circunference of reference, so one has not anymore a
natural time coordinisation between the clocks at different heights. But,
as we have seen in b), such a coordinisation is unavoidable for a
(global) characterization of the trajectory (besides, the drift is also a
non-local effect to be measured). So the
observers at different heights made the gentlemen agreement to pace their
clocks by the clock at the reference level and, with this agreement, what
their devise will measure will be $\vec{a}_{FC}$ (which differs from
$\vec{a}$ by $\vec{g}\cdot\vec{v}_{FC}\vec{v}_{FC}$) and a non-null drift
of the circular horizontal projection of the particle trajectory.

\end{description}

Note that if one tries to reconstruct the theoretical scheme of Moliner
et al experiment (that is, equations (\ref{40})-(\ref{42}) etc) by using
directly $\vec{a}$ instead of $\vec{a}_{FC}$, then the term
$\vec{g}\cdot\vec{v}_{FC}\;\vec{v}_{FC}$ automatically desappears from
(\ref{42}) and the theoretical scheme becomes vacuous. So the scheme is
consistent just with the empirical, observable, content referred above in
a) and b).

In face of this fact that Moliner et al experiment does not test directly
the GEM expression or Mishra's result, we must reexamine to what extent
the physical content effectively involved in the test really includes them.

The following reasoning can be done:

\begin{description}

\item a) in the general GEM formula, and in its canonical representation,
$a_{fw}$ (resp. $\vec{a}$) is a sum of two parcels, one according to
$\hat{g}$ and the other according to $\hat{v}$. Its reduction to Mishra's
expression comes for $\hat{v}=\hat{g}$, ie, is made up of the
contributions of both these parcels. So, any experiment able to verify
separately each parcel in $a_{fw}$ (resp. $\vec{a}$) confirms the entire
general formula and, in particular, Mishra's expression;

\item b) the comparison between $\vec{a}$ and $\vec{a}_{FC}$ ((\ref{30})
and (\ref{39})) shows that each parcel of $a_{fw}$ (resp. $\vec{a}$)
corresponds to the respective parcel of $\vec{a}_{FC}$. So, any
experiment able to verify separately both parcels
of $\vec{a}_{FC}$ implies the same kind of verification for $a_{fw}$
(resp. $\vec{a}$) (and reciprocally) and so confirms the GEM expression
and, consequently, Mishra's equation;

\item c) fortunately, the Moliner et al experiment tests, separately,
each parcel of $\vec{a}_{FC}$: the measurement of the electric field
necessary to avoid the charge falling down during the circular movement
tests the parcel according to $\hat{g}$; and
the measurement of the horizontal drift velocity tests the parcel
according to $\hat{v}$.

\end{description}

Thus we can conclude that, if the experiment gives both the predicted
electric field and the forseen drift velocity, it will confirm the
generic static GEM expressions for the general free fall and, in
particular, Mishra's one-dimensional result. Besides,
it will favour the soundness of Vigier's solution on the nature of
inertial forces.

\section*{Acknowledgements}

We would like to thank Drs. M. Portilla and B.M. Pimentel for interesting
discussions. We are also indebted to the anonimous referees for their
contributive suggestions.

\end{document}